\documentclass[referee]{cjaa}           

\usepackage{graphicx}                   
\input{epsf.sty}                        
\input{psfig.sty}                       

\begin{document}

   \title{Physical Limits of Different Models of Cosmic Gamma-Ray
   Bursts
}


   \author{Gennady S. Bisnovatyi-Kogan
      \inst{1,2}\mailto{}
      }
   \offprints{G.S. Bisnovatyi-Kogan}                   

   \institute{Space Research Inst. Rus. Acad. Sci, Profsoyuznaya
84/32, Moscow 117997, Russia\\
\email{gkogan@mx.iki.rssi.ru}
        \and
Joint Institute of Nuclear Researches, Dubna, Russia
             \\
  }

   \date{Received~~2003 month day; accepted~~2003~~month day}

   \abstract{
The present common view about GRB origin is related to cosmology,
what is based on statistical analysis, and on measurements of the
redshifts in the GRB optical afterglows of long GRB. No
correlation is found between redshifts, GRB spectrum, and total
GRB fluence. Comparison of KONUS and BATSE data about statistics
and hard X-ray lines is done, and some differences are noted. Hard
gamma-ray afterglows, prompt optical spectra, hard X-ray lines
measurements could be important for farther
insight into GRB origin. Possible possible connection of short GRB
with soft gamma repeaters is discussed.
   }

   \authorrunning{G.S. Bisnovatyi-Kogan}            
   \titlerunning{Physical limits of gamma-ray bursts models
   }  


   \maketitle
%
%
\section{Introduction}           
\label{sect:intro}
It is generally accepted now that cosmic gamma-ray bursts (GRB)
have a cosmological origin. The first cosmological model, based on
explosions in active galactic nuclei (AGN) was suggested by
Prilutsky and Usov (\cite{pu}). A mechanism of the GRB origin in
the vicinity of a collapsing object based on neutrino-antineutrino
annihilation was analyzed by Berezinsky and Prilutsky (\cite{bp}).
GRB production in
supernova explosion was suggested by Bisnovatyi-Kogan et al. (\cite{bkin}).
Here we discuss
different observational features of GRB, analyze difficulties and
problems of their interpretation in the cosmological model, and
physical restrictions to their model. At the end we are analyze a
possible connection between short GRB and soft gamma repeater
(SGR).

\section{GRB physical models}
\label{sec:2}

The GRB models may be classified by two levels. The upper one is
related directly to the observational appearance, and include 3
main models.

{\bf 1}. Fireball. {\bf 2}. Cannon ball (or gun bullet). {\bf 3}.
Precessing jets.

\noindent The main restrictions are connected with the next
(basic) level of GRB model, which is related to energy source,
producing a huge energy output necessary for a cosmological GRB
model. These class
 contains 5 main models.

{\bf 1}. (NS+NS), (NS+BH) mergers. This mechanism was investigated
numerically by Ruffert and Janka (\cite{rj98,rj99}).
Gamma radiation is produced here
by $(\nu, \tilde{\nu})$ annihilation, and the energy output is not
enough to explain most powerful GRB even with account of strong
beaming. The energy emitted in the isotropic optical afterglow of
GRB 990123 (Akerlof et al., \cite{ak2}; Kulkarni et al. \cite{ku})
is about an order of magnitude larger
than the total radiation energy output in this model.

{\bf 2}. Magnetorotational explosion.
 Magnetorotational explosion, proposed by Paczynkki (\cite{pa}) for
a cosmological GRB, had been suggested earlier for the supernova
explosion by Bisnovatyi-Kogan (\cite{bk}).
Numerical calculations gave the efficiency
of a transformation of the rotational energy into the kinetic one
at the level of few percent (Ardelyan et al., \cite{ard,abm}).
This is enough for the
supernovae energy output, but is too low for cosmological GRB.

{\bf 3}. Hypernova.
 This model denoting very powerful supernova,
was suggested in general by Paczynski (\cite{pa}), and is popular now
because traces of the supernova explosions are believed to be
found in the optical afterglows of several GRB (Sokolov, \cite{sok};
Dado et al., \cite{ddd}; Stanek et al., \cite{sta}).
The action of strongly magnetized rapidly rotating new born
neutron star for production of GRB was considered by Usov (\cite{us}).
Another
hypernova model is based on a collapse of a massive core,
formation of a black hole $M_{bh}\sim 20 M_{\odot}$, surrounded by
a massive disk with a rapid accretion and appearance of GRB (MacFadyen
and Woosley, \cite{mwoo}). This modes seems to be most promising now.

{\bf 4}. Magnetized disks around rotating (Kerr) black holes
(RBH).
 This model is based on extraction of rotating energy of RBH when
magnetic field is connecting the RBH with the surrounding
accretion disk or accretion torus (van Putten, \cite{vpu}).

{\bf 5}. In the model proposed Ruffini et al. (\cite{rswx})
GRB is created by
the pair-electromagnetic pulse from an electrically charged black
hole surrounded by a baryonic remnant. The main problem here is
how to form such a strongly charged BH.

\section{Basic observational data}
\label{sec:3}
{\bf\normalsize 3.1. Statistics}

Statistical arguments in favor of the cosmological
origin of GRB are based on a visual isotropy of GRB distribution
on the sky in combination with a strong deviation of $\log N\, -\,
\log S$ distribution obtained in BATSE observations
(Meegan et al., \cite{mee})
 from the euclidian uniform distribution with the
slope $3/2$. Similar properties have been obtained in KONUS
experiment (Mazets et al., \cite{kon80}) where the authors explained deviations
from $3/2$ slope by selection effects. The analysis of KONUS data
with account of selection effects made by Higdon and Schmidt (\cite{hs90}) gave the
average value $<V/V_{max}>=0.45 \pm 0.03$; the value 0.5
corresponds to pure uniform distribution. KONUS data had been
obtained in conditions of constant background. Similar analysis
made by Schmidt (\cite{s99}) of BATSE data, obtained in conditions of substantially
variable background, gave resulting $<V/V_{max}>=0.334 \pm 0.008$.
These two results seems to be in contradiction, because KONUS
sensitivity was only 3 times less than that of BATSE, where
deviations from the uniform distribution in BATSE data are sill
large (Fishman and Meegan, \cite{fm95}).

\begin{figure}[h]
\centering
\includegraphics
{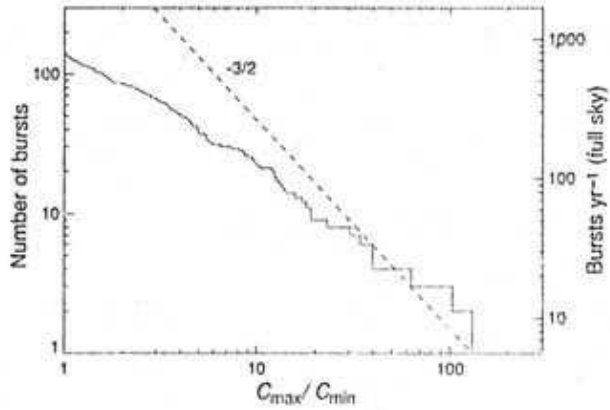}
\caption[h]{Integral number distribution of 140 GRB as a
function of peak rate . A -3/2 power law is expected for a
homogeneous distribution of sources. The full sky rate is $\sim
800$ bursts per year, from Meegan et al. (\cite{mee}).}
\label{batse}
\end{figure}

\begin{figure}
   \plottwo{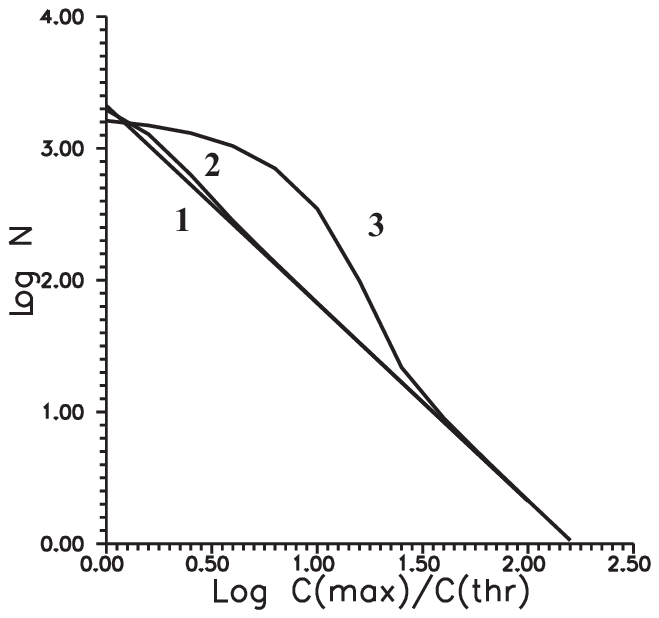} {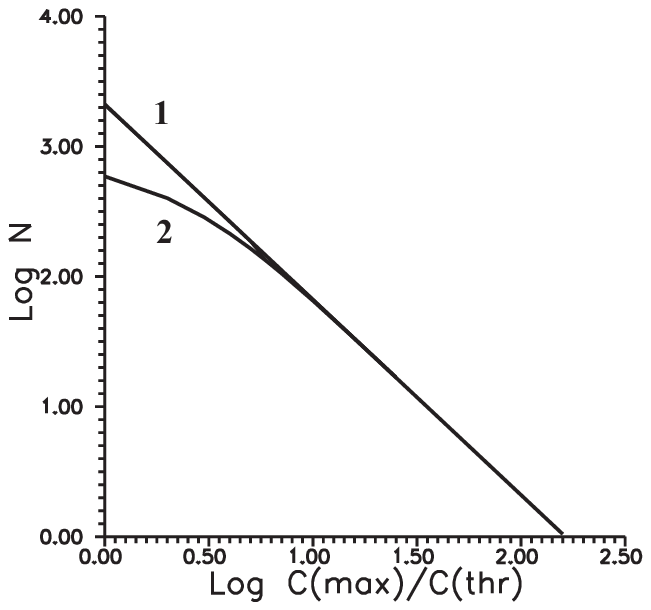}
   \caption{{\bf a.} The curve $[\log N - \log C(max)/C(thr)]$ in presence of
stochastic errors, distributed according to normal
distribution with an average error $\Delta_1$ in units of a
threshold;
1 - straight line with slope 3/2, corresponding
to $\Delta_1=0$; 2 - curve with $\Delta_1=1$;
3 - curve with $\Delta_1=10$.
{\bf b.} Same as in {\bf a} for normal logarithmic distribution,
$\Delta$ determines a logarithm of number of thresholds, as an
average error;
 1 - straight line with a slope 3/2, corresponding
to $\Delta=0$; 2 - curve with $\Delta=1$; $C(max)$ is the peak intensity
of the burst; $C(thr)$ is a corresponding threshold value,
from Bisnovatyi-Kogan (\cite{bk3}).}
   \label{stat}
\end{figure}

Detailed statistical analysis and calculation of of BATSE data,
divided in 4 classes according to their hardness and calculation
of  $<V/V_{max}>$ for different classes have been done by
Schmidt (\cite{s01}). In the cosmological model we may expect
smaller value of $<V/V_{max}>$ for softer GRB in the case of a
uniform sample, because larger red shifts would correspond to
softer spectra. The result is quite opposite, and soft GRB have
larger  $<V/V_{max}>$ than the hard ones, 0.47 and 0.27
respectively. It is supposed by Schmidt (\cite{s01})
that strong excess of
luminosity in hard GRB overcomes the tendency of the
uniform sample. The possibility of decisive role of selection
effects (incompleteness of data, statistical errors in estimation
of luminosity in presence of the threshold) was analyzed by
Harrison et al. (\cite{har95}), Bisnovatyi-Kogan (\cite{bk3}).
 The influence of statistical errors in presence of the
threshold was analyzed by Bisnovatyi-Kogan (\cite{bk3}).
The  $\log N\, -\, \log S$
curve in presence of statistical errors on the level of average 10
thresholds has a similarity with the BATSE distribution
(see Fig~\ref{batse} and Fig~\ref{stat}).

{\bf\normalsize 3.2. Optical Afterglows and Red Shifted Lines}

The spectra of optical afterglows have shown large red shifts $z$,
up to 4.5, indicating to the cosmological origin of GRB and their
enormous energy outputs. In most cases the red shifts have been
measured in the faint host galaxies. The list of red shift
measurements is given in the Table~\ref{tab1},
where data about redshifts from Djorgovski et al. (\cite{kul01})
are completed by total GRB fluences (Bisnovatyi-Kogan, \cite{bkvul}). This
table contains the trigger number and fluence from 4B catalogue
(Paciesas et al., \cite{cat4}), and fluence for the GRB from other references. Huge
energy output during a short time (0.1 - few 100 seconds) create
problems for the cosmological interpretation.

\begin{figure}[h]
\centering
\includegraphics[height=9cm]{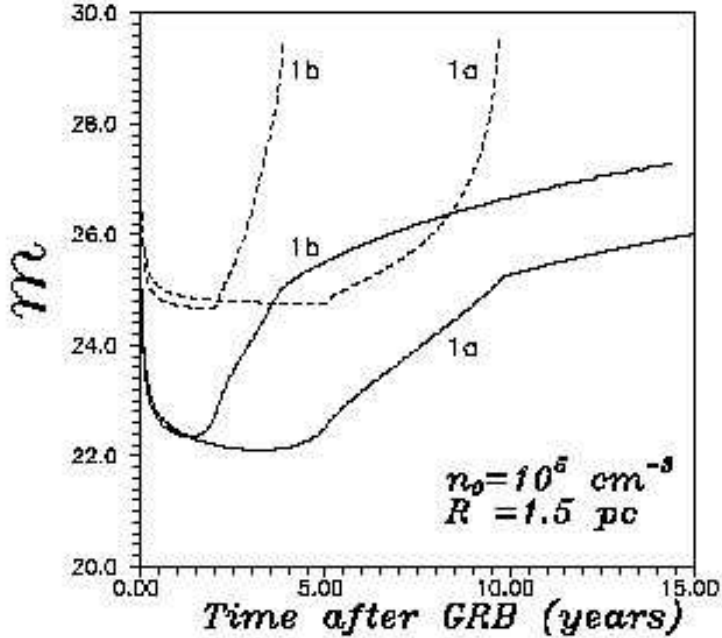}
\caption[h]{
The magnitudes of the counterparts (upper limit - solid line,
 lower limit - dashed line) as a function of time after burst
for GRB with total flux near the Earth $F_{\rm GRB} = 10^{-4}$ erg
cm$^{-2}$:
1a. - for the case E = $10^{52}$ erg; $n_0= 10^5$ cm$^{-3}$;
1b -  for the case E = $10^{51}$ erg;  $n_0= 10^5$ cm$^{-3}$, from
Bisnovatyi-Kogan and Timokhin (\cite{bkt}).}
\label{bktim}
\end{figure}

It was shown first by Paczynski (\cite{pa}), that properties of GRB afterglows
are explained better under suggestion that GRB source is situated
in a dusty star forming region with a high gas density.
Interaction of mighty GRB pulse with the surrounding gas with
a density $n=10^4\,-\,10^5$ cm$^{-3}$ create a specific form of the
optical afterglow, lasting up to few tens of years. The
calculations of light curve and spectrum of such afterglow
have been done by Bisnovatyi-Kogan and Timokhin (\cite{bkt}).
Some results are represented in
Fig.~\ref{bktim}.
It was shown, that counterparts of cosmological GRB due to
interaction of gamma-radiation with dense interstellar media
are "long-living" objects, existing for years after GRB.
To distinguish GRB counterpart from a supernova event,
having similar energy output, it is necessary to take into
account its unusual light curve and spectrum. In the optical
region of the spectrum the strongest emission lines are
$H_{\alpha}$ and $H_{\beta}$.
Discovery of even one optical counterpart of
GRB with properties described above would give an opportunity
to probe the density of the interstellar medium
around the burst, and therefore would give an indication
of the burst progenitor.

{\bf\normalsize 3.3. Collimation}

To avoid a huge energy production strong collimation is suggested
in the radiation of GRB. In the "cannon-ball" model of
Dado et al. (\cite{ddd}) the
bulk motion Lorentz factor is $\Gamma \approx 10^2\,-\,10^3$,
leading to collimation factor $\Omega \approx
10^{-4}\,-\,10^{-6}$. Analysis of GRB collimation have been done
by Rhoads (\cite{rho}). The main restriction to the collimation angle
follows from the analysis of the probability of appearance of the
orphan optical afterglow, which most probably have low or no
collimation. The absence of any variable orphan afterglow in a
search poses the following restrictions. It was expected to detect
$\sim 0.2$ afrerglows, if bursts are isotropic, so the absence of
orphan afterglows suggests $\Omega_{opt}/\Omega_{\gamma}<<½ 100$,
which is enough to rule out the most extreme collimation
scenarios. At radio wavelengths published source counts and
variability studies have been used by Perna and Loeb (\cite{pl98})
to place a limit
on the collimation angle, $\theta_{\gamma} \geq 5^{\circ}$.
Because radio afterglows last into the non-relativistic phase of
the GRB remnant evolution, the radio afterglows are expected to
radiate essentially isotropically, and the orphan afterglow limits
on radio $\Omega_r/\Omega_{\gamma}$ immediately imply a limit on
$\Omega_{\gamma}$ itself.

\begin{table}
\centering \caption{GRB Host Galaxies, Redshifts and Fluences
(June 2001)} \label{tab1}
\begin{tabular}{|l|l|l|l|l|l|l|} \hline\noalign{\smallskip}
 Trigger&
  GRB &
  $R$ mag &
  Redshift &
  Type $^a$&
  Fluence$^e$\\
number&&&& & erg/cm$^2$\\
\noalign{\smallskip} \hline \noalign{\smallskip}
& 970228     &   25.2 &  0.695   & e & 10$^{-5}$  \\
6225& 970508     &   25.7 &  0.835   & a,e & $3.5\cdot 10^{-6}$(3+4) \\
6350& 970828     &   24.5 &  0.9579  & e & $7 \cdot 10^{-5}$ \\
6533& 971214     &   25.6 &  3.418   & e & $ 10^{-5}$(3+4)  \\
6659& 980326     &   29.2 &$\sim$1?  &   & $6.3 \cdot 10^{-7}$(3+4) \\
6665& 980329     &   27.7 &$<$3.9    & (b)&$7.1\cdot 10^{-5}$(3+4)\\
6707& 980425 $^c$&   14   &  0.0085  & a,e&$4.4 \cdot 10^{-6}$  \\
6764& 980519     &   26.2 &          &  &$9.4 \cdot 10^{-6}$(all 4)     \\
& 980613     &   24.0 &  1.097   & e &$1.7\cdot 10^{-6}$ \\
6891& 980703     &   22.6 &  0.966   & a,e&$5.4\cdot 10^{-5}$(3+4) \\
7281& 981226     &   24.8 &          & &$2.3\cdot 10^{-6}$(3+4)   \\
7343& 990123     &   23.9 &  1.600   & a,e &$5.1 \cdot 10^{-4}$\\
7457& 990308 $^d$&$>$28.5 &          &    &$1.9\cdot 10^{-5}(3+4)$ \\
7549& 990506     &   24.8 &  1.30    & e  &$2.2 \cdot 10^{-4}$  \\
7560& 990510     &   28.5 &  1.619   & a &$2.6 \cdot 10^{-5}$ \\
& 990705     &   22.8 &  0.86    & x&$\sim 3 \cdot 10^{-5}$     \\
& 990712     &   21.8 &  0.4331  & a,e& \\
& 991208     &   24.4 &  0.7055  & e&$\sim 10^{-4}$  \\
7906& 991216     &   24.85&  1.02    & a,x&$2.1\cdot 10^{-4}$(3+4) \\
7975& 000131     &$>$25.7 &  4.50    & b &$\sim 10^{-5}$  \\
& 000214     &        &0.37--0.47& x&$\sim 2 \cdot 10^{-5}$    \\
& 000301C    &   28.0 &  2.0335  & a  & $\sim 4\cdot 10^{-6}$ \\
& 000418     &   23.9 &  1.1185  & e&$1.3\cdot 10^{-5}$  \\
& 000630     &   26.7 &          &   &$2\cdot 10^{-6}$  \\
& 000911     &   25.0 &  1.0585  & e &$5 \cdot 10^{-6}$  \\
& 000926     &   23.9 &  2.0369  & a &$2.2 \cdot 10^{-5}$  \\
& 010222     &$>$24   &  1.477   & a& brightest of   \\
&&&&&BeppoSAX\\
\hline
\end{tabular}
\\
{Notes}:
$^a$ e = line emission, a = absorption, b = continuum break, x = x-ray \\
$^c$ Association of this galaxy/SN/GRB is somewhat controversial \\
$^d$ Association of the OT with this GRB may be uncertain \\
$^e$The number of BATSE peak channel is indicated in brackets, from \cite{cat4}, \\
otherwise the estimation of bolometric fluence from other sources,
made in \cite{bkvul} is indicated \\
\medskip
\hrule
\vspace*{-13pt}
\end{table}

\begin{figure}
\centering
\includegraphics[height=8cm]{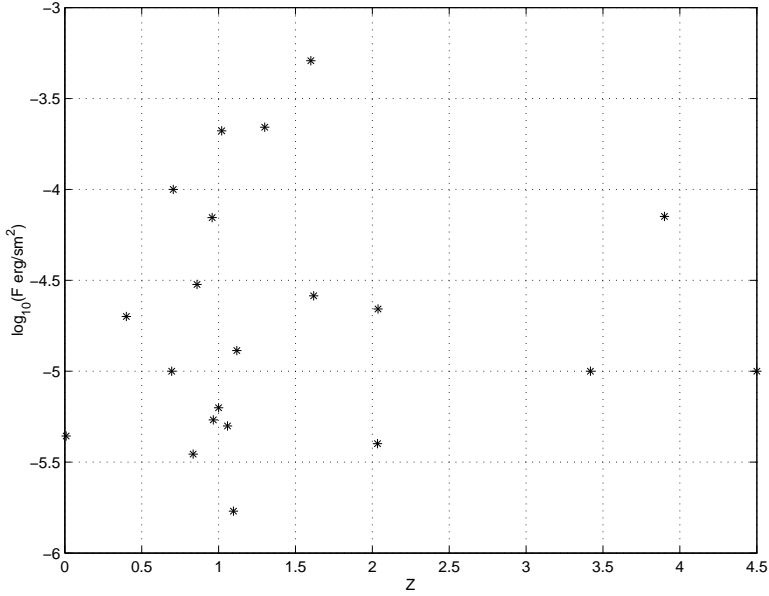}
\caption[h]{Fluence $F$ versus redshift $z$ for GRB from the Table
1. }
\label{bkz}
\end{figure}

Comparison of the red shifts and fluences from Table~\ref{tab1} shows no
correlation between distance and observed flux (see Fig~\ref{bkz}). It is
explained by strong collimation, and strong scattering is
connected with different sight angles in the beam. If the
collimation is connected with the relativistic bulk motion
(Dado et al., \cite{ddd}), than strong correlation is expected
between GRB duration and their
power: stronger GRB should be shorter. Absence of such correlation
excludes models based on the relativistic bulk motion collimation.

{\bf\normalsize 3.4. Prompt Optical Afterglows}

\begin{figure}[h]
\centering
\includegraphics[height=8cm]{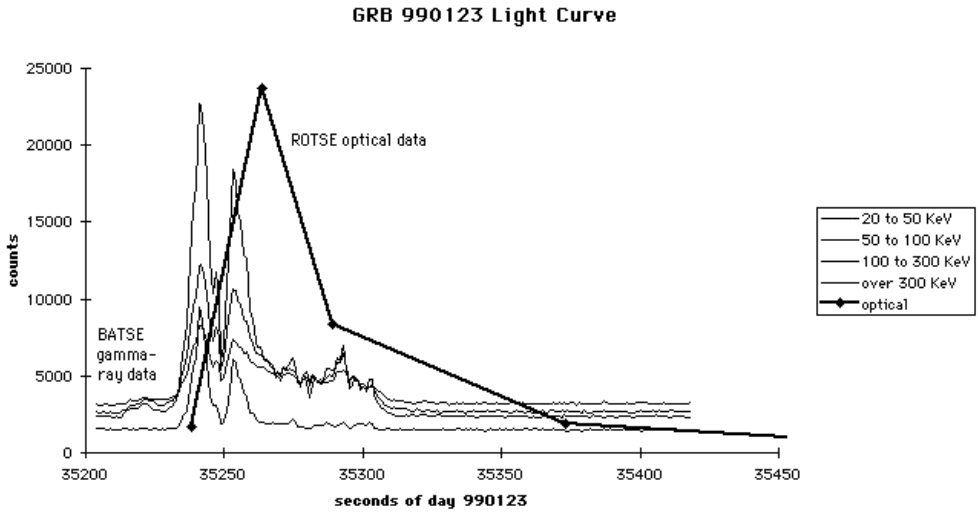}
\caption[h]{
Thin lines represent BATSE gamma-ray profile with 1024 ms resolution
in different energy channels. The thick line represents the first few
frames of ROTSE data. ROTSE began taking data 22 seconds after the initial
trigger. This is the brightest
ever optical burst, from Pozanenko, (\cite{poz}).}
\label{opt99}
\end{figure}

\begin{figure}[h]
\centering
\includegraphics[height=4.5cm]{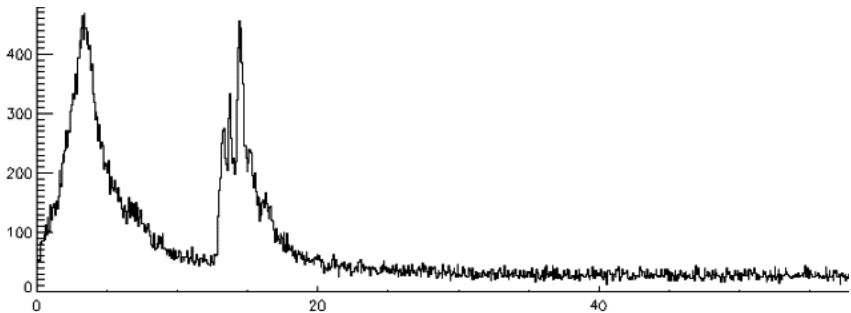}
\caption[h]{GRB030329 in gamma-ray:
 Konus-Wind (UT)11:37:29.
Fluence = $1.2 \times 10^{-4}$ erg /cm$^2$,
Duration = ~50 s,
Peak Flux = $2.5 \times 10^{-5}$ erg /cm$^2$ / s;
One of the most luminous burst (from ~4000).
 From Pozanenko, (\cite{poz}).}
\label{gam03}
\end{figure}

\begin{figure}[h]
\centering
\includegraphics[height=8cm]{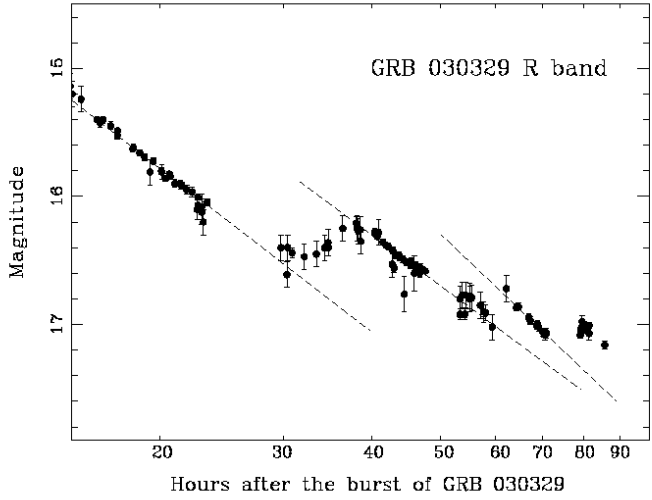}
\caption[h]{Variability in hours scale
of the optical afterglow of GRB030329,
 from Pozanenko, (\cite{poz}).}
\label{opt03}
\end{figure}

The afterglow of
GRB 990123 was catched by optical observations 22 seconds
after the onset of the burst (Akerlof et al., \cite{ak1,ak2}).
 GRB 990123 was
detected by BATSE on 1999 January 23.407594. The event was strong
and consisted of a multi-peaked temporal structure lasting
$\ge$100 s
(see Fig.~\ref{opt99}), with significant spectral evolution. The T50 and T90
durations are 29.82 ($\pm$ 0.10) s and 63.30 ($\pm$ 0.26) s,
respectively. The maximum optical brightness 8.95$^m$ was reached
30 sec. after the GRB beginning, and after 95 sec. it was already
of 14.5$^m$. So the gamma ray maximum almost coincides with the
optical one. The observed optical luminosity, related to the red
shift $z=1.61$ reaches $L_{opt} \approx 4 \cdot 10^{49}$ ergs/s,
what is about 5 orders of magnitude brighter than optical
luminosity of any observed supernova. The energy of the prompt
optical emission reaches $10^{51}$ ergs, and the isotropic
gamma-ray flux is about $2.3 \cdot 10^{54}$ ergs, what exceeds the
rest energy of the Sun (Akerlof et al., \cite{ak2};
Kulkarni et al., \cite{ku}).
Another bright afterglows
have been observed in GRB 021004  (15m, z=2.3), GRB 030329
(12.4m, z=0.168) and GRB 030418  (16.9m). Brightest visual
magnitude and redshift are given in brackets. The most remarkable
afterglow observed by many observatories was in GRB 030329 (see
i.g. Rumjantsev et al., \cite{poza}; Burenin et al., \cite{bur}),
where supernova was probably detected by the
features of spectra (Stanek et al., \cite{sta}).
Konus-Wind observations of GRB 030329
are given in Fig.~\ref{gam03}, and light curve of the optical afterglow
obtained in Crimea observatory is represented in Fig.~\ref{opt03}
(from Pozanenko, \cite{poz}).

 {\bf\normalsize 3.5. High-Energy Afterglow}

EGRET observations on CGRO have shown that GRB emit also very hard
gamma photons up to 20 GeV (Fishman and Meegan, \cite{fm95}).
The number of GRB with
detected hard gamma radiation is about 10, from them 5 bursts had
registered photon energies over 100 MeV
(Schneid et al., \cite{schneid}). Hard gamma
emission continues up to 1.5 hours in the GRB940217. Comparison of the
angular aperture of EGRET and BATSE leads to conclusion that hard
gamma radiation could be observed in large fraction (about one
half) of all GRB. Spectral slope in hard gamma region lays between
(-2) and (-3.7), and varies rapidly, becoming softer with time
(GRB920622 in Schneid et al. (\cite{schneid2})).
 Data about spectra of hard gamma radiation of
radio pulsars in Crab nebula (Much et al., \cite{crab}),
and PSR B1055-52 (Thompson et al., \cite{gpsr})
show similar numbers and variety. With account of non-pulsed
Crab spectrum the slope varies between (-1.78) and (-2.75). If GRB
is connected with SN explosion and neutron star formation,
than residual oscillations of the neutron star
may be responsible for the extended hard gamma ray
afterglow (Bisnovatyi-Kogan, \cite{bk95}; Timokhin et al., \cite{tbk},
Ding and Cheng, \cite{ding}).
 .

{\bf\normalsize 3.6. Hard X-Ray Lines}

\begin{figure}[h]
\centering
\includegraphics[height=7cm]{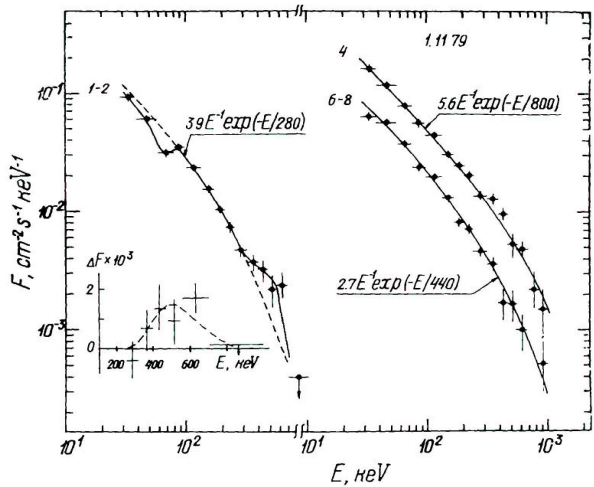}
\caption[h]{Spectral evolution of 1 November 1979 burst. (1-2) - spectrum
obtained in the first 8 s with absorption line at $\approx 65$ keV and broad
emission feature at 350-650 keV; (4) - spectruim measured in the 4-th 4 s
interval; (6-8) - spectrum summed over 6-th, 7-th and 8-th intervals,
from Mazets et al. (\cite{ma82}).}
\label{fig8ma}
\end{figure}

\begin{figure}
   \plottwo{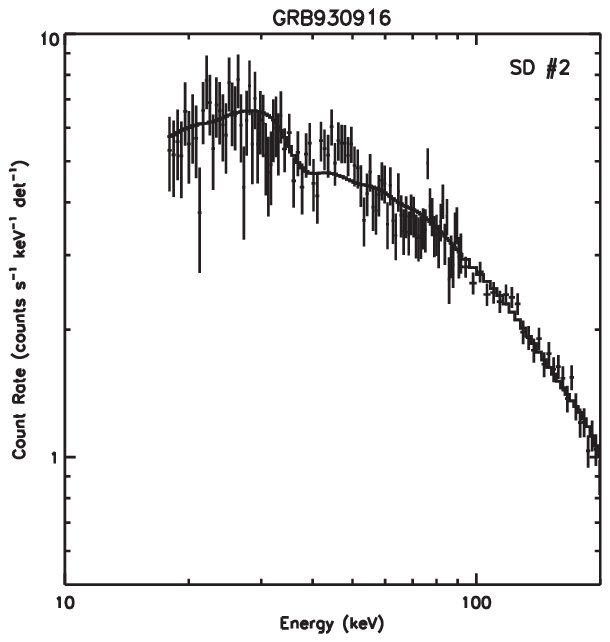} {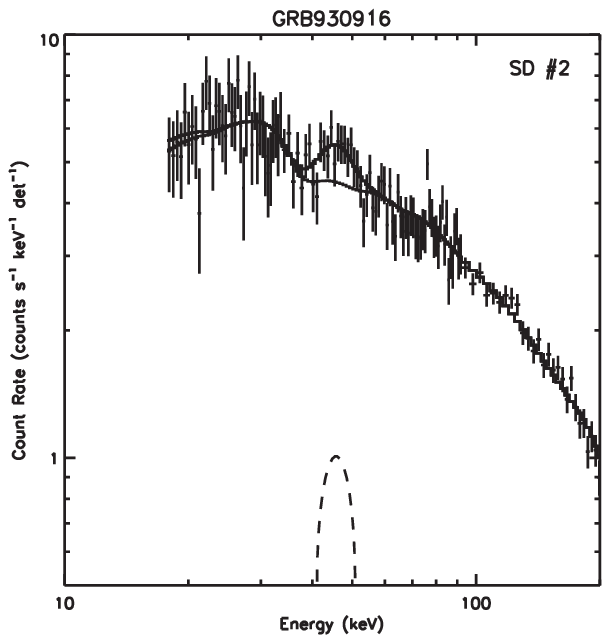}
   \caption{Data from the interval 22.144 to 83.200~s after the BATSE trigger
of GRB930916.
The `bump' at 30 keV is expected from the K-edge of the iodine in NaI.
Left panel: best continuum-only fit to the data of SD~2.
Right panel: A narrow spectral feature is added to the model: an emission line
at 45 keV improves $\chi^2$ by 23.1.
The solid histogram depicts the total count model; the
dashed histograms show the continuum and line portions separately,
from  Briggs et al. (\cite{briggs}).
}
   \label{batlines}
\end{figure}

Hard gamma-ray lines in GRB spectra have been discovered by KONUS
group (Mazets et al.,\cite{kon82}).
They had been interpreted there as cyclotron
lines, and have been seen in 20-30\% of the GRB. These spectra had
shown a distinct variability: the visible absorption decreases
with time (Fig.~\ref{fig8ma}).
In BATSE data existence of hard
X-ray spectral features in GRB spectra was found by Briggs et al.
(\cite{briggs}). In this paper 13
statistically significant line candidates have been found from 117
investigated GRBs.
One of the best cases for detecting a line is GRB~941017 in
which the data from two detectors are consistent. In some GRBs the
line was found only by one detector, while it was not
statistically significant in the other one. The data
for GRB930916 from the detector with the well observed line
is given in Fig.~\ref{batlines} from
Briggs et al. (\cite{briggs}).
The conclusion
of this paper is that the reality of all of the BATSE line
candidates is unclear. Note however that spectra of GRB930916
 had been obtained 20 s after the
trigger, and according to Mazets et al. (\cite{kon82})
the lines are the strongest
at the beginning of the burst (see Fig.~\ref{fig8ma}).
The only interpretation in the cosmological model
(Hailey et al., \cite{grblines})
is based on the
blue-shifted ($\Gamma=25-100$) spectrum of the gas cloud
illuminated by the gamma radiation of the fireball. Similar model
was suggested by Bisnovatyi-Kogan and Illarionov (\cite{bki89})
for explanation of the lines
observed by KONUS.

{\bf\normalsize 3.7. GRB and Supernovae}

The following GRB have evidences for their connection with SN:
GRB980425 (z=0.0085, 40 Mpc),
GRB980326 (z=1),
GRB011121 (z=0.365),
GRB020405 (z=0.695),
GRB030329 (z=0.169).
Some optical afterglows have a "red bump" during 15-75 days, what may be
consistent with the underlying SN explosion,
see i.g. Sokolov (\cite{sok}). Nevertheless the
SN-GRB connection is not quite certain.

 \section{Short GRB and SGR}

The statistical analysis
reveals at least two separate samples consisting of long ($>\sim
2$~s) and short bursts. Optical afterglows and redshift
measurements have been done only for long bursts. Therefore, it is
not excluded that short bursts have different (may be galactic)
origin. Compare properties of short GRB
with giant bursts from soft gamma-repeaters (SGR) inside the
Galaxy. From the larger distance only giant bursts would be
registered, which could be attributed to short GRB.  The
existence of giant bursts in the SGR (3 in 4 firmly known SGR
in the Galaxy and LMC) implies a possibility for observation of
giant bursts, which appear as short GRB, in other neighboring
galaxies. The estimation gives more than 10 expected "short GRB"
of this type from M 31 and other close neighbors.
The absence of any GRB projecting on the local group galaxies may
indicate that SGR are more close and less luminous objects, than
it is now accepted  (Bisnovatyi-Kogan, \cite{bkvul99}).

\section{Conclusions}
There is no fully consistent GRB model: neither radiation, nor explosion.
It is not excluded, that short GRB have galactic origin, and
giant bursts in SGR are connected with short GRB (Mazets et al., \cite{ma82}).
Critical experiments are needed: spectra of prompt optical afterglows;
study of hard gamma-ray afterglows; search for orphans optical afterglows in
optical all sky monitoring.
Cosmological GRB may come from
collapse of massive rotating star followed by
formation of Kerr black hole surrounded by massive magnetized disk,
and rapid accretion leading to GRB; or from some exotic models.

\begin{acknowledgements}
The author is grateful to F.Giovannelli and the organizers of
the workshop for support and hospitality. Author acknowledges
partial support from RFBR grant 02-02-16900 and INTAS grant 00491.
\end{acknowledgements}

\label{lastpage}

\end{document}